\documentclass[11pt]{article}
\usepackage{graphicx}

\newcommand\pubnumber{~}
\newcommand\pubdate{\today}

\def\napoli{
Institute of High Energy Physics, CAS, Beijing, China}
\def\support{\footnote{
Work supported in part by the National Natural Science Foundation of China (NSFC) under Contract No.10935007. 
I would like to thank Yi Fang and Hailong Ma for many helpful discussions about this experimental review. 
}}

\def\Title#1{\begin{center} {\Large #1 } \end{center}}
\def\Author#1{\begin{center}{ \sc #1} \end{center}}
\def\Address#1{\begin{center}{ \it #1} \end{center}}

\newcommand\pubblock{\rightline{\begin{tabular}{l} \pubnumber\\
         \pubdate  \end{tabular}}}
\newenvironment{Abstract}{\begin{quotation}  }{\end{quotation}}
\newenvironment{Presented}{\begin{quotation} \begin{center} 
             PRESENTED AT\end{center}\bigskip 
      \begin{center}\begin{large}}{\end{large}\end{center} \end{quotation}}

\def\beq{\begin{equation}}
\def\eeq#1{\label{#1}\end{equation}}
\def\eeqn{\end{equation}}

\def\beqa{\begin{eqnarray}}
\def\eeqa#1{\label{#1}\end{eqnarray}}
\def\eeqan{\end{eqnarray}}

\let\bar=\overbar

\def\Dslash{\not{\hbox{\kern-4pt $D$}}}
\def\dslash{\not{\hbox{\kern-2pt $\del$}}}

\def\msb{{\bar{\ssstyle M \kern -1pt S}}}

\begin{document}
\begin{titlepage}
\pubblock

\vfill
\Title{
Recent Experimental Results on Leptonic $D^+_{(s)}$ Decays, Semileptonic $D$ Decays and Extraction of $|V_{cd(s)}|$}
\vfill
\Author{ Gang Rong\support}
\Address{\napoli}
\vfill
\begin{Abstract}
The recent experimental results on leptonic $D^+_{(s)}$ decays, semileptonic $D$ decays, determinations of
decay constants and form factors, as well as extractions of $|V_{cd}|$ and $|V_{cs}|$
are briefly reviewed. 
Global analysis of all existing measurements of leptonic $D^+_{(s)}$ decays and semileptonic $D$ decays
yields $|V_{cd}|=0.2157\pm 0.0045$ and $|V_{cs}|=0.983\pm 0.011$, which are the most precision determinations
of these two CKM matrix elements up to date.
\end{Abstract}

\vfill
\begin{Presented}
The 8th International Workshop on the CKM Unitarity Triangle (CKM 2014), \\
Vienna, Austria, September 8-12, 2014
\end{Presented}
\vfill
\end{titlepage}
\def\thefootnote{\fnsymbol{footnote}}
\setcounter{footnote}{0}

\section{Introduction}

In the Standard Model (SM) of particle physics, 
the decay rate of $D^+_{(s)}\to\ell^+\nu_\ell$ (where $\ell = e$, $\mu$, or $\tau$) 
relates to a product of the decay constant
$f_{D^+_{(s)}}$ and the Cabibbo-Kobayashi-Maskawa (CKM)
matrix  element $V_{cd(s)}$ by
\begin{equation}
\Gamma(D^+_{(s)} \rightarrow \ell^+\nu_{\ell})=
     \frac{G^2_F } {8\pi}
      m^2_{\ell} m_{D^+_{(s)}}
    \left (1- \frac{m^2_{\ell} } {m^2_{D^+_{(s)}}}\right )^2 f^2_{D^+_{(s)}}\mid V_{cd(s)} \mid^2 ,
\label{eq01}
\end{equation}
where $G_F$ is the Fermi coupling constant,
$m_{\ell}$ is the mass of the lepton, and $m_{D^+_{(s)}}$ is the mass of the $D^+_{(s)}$ meson.
The differential rate of
$D \to \pi(K) e^+\nu_e$ decay relates to a product of decay form factor $f_+^{\pi(K)}(q^2)$ and
$V_{cd(s)}$ by
\begin{equation}
\frac {d\Gamma }{dq^2} = X \frac {G_F^2}{24\pi ^3}
|\vec p_{\pi(K)}|^3
|f_+^{\pi(K)}(q^2)|^2 |V_{cd(s)}|^2,
\label{eq_dGamma_dq2}
\end{equation}
where $q^2$ is square of the four-momenta transfer,
$\vec p_{\pi(K)}$ is the three-momentum of the
$\pi$ ($K$) meson in the rest frame of the $D$ meson,
and $X$ is a factor due to isospin, which equals to $1$ for $D^0\to\pi^-e^+\nu_e$, $D^0\to K^-e^+\nu_e$
and $D^+\to\bar K^0e^+\nu_e$, and equals to $1/2$ for $D^+\to\pi^0e^+\nu_e$.

With precision measurements of these decay rates one can more precisely determine 
the decay constants $f_{D^+_{(s)}}$ and form factors $f^{\pi(K)}_+(0)$ as well as these CKM matrix elements. 
The precisely measured values of 
$f_{D^+_{(s)}}$ and $f^{\pi(K)}_+(0)$ 
can be used to validate LQCD calculations for
these quantities, which can improve determinations of the CKM matrix elements $|V_{ub}|$, $|V_{td}|$, and $|V_{ts}|$.
As a consequence, the uncertainty in the overall constraint on the unitarity triangle of the CKM matrix can be reduced.
These can be used for more stringent test of the SM and search for New Physics beyond the SM.

\section{Results on $D^+_{(s)}\rightarrow \ell^+\nu_{\ell}$ decays and decay constants $f_{D^+_{(s)}}$}
\subsection{New results on $D^+_{(s)}\rightarrow \ell^+\nu_{\ell}$ decays}
    In 2014, the BESIII Collaboration made a measurement of 
$D^+ \rightarrow \mu^+\nu_\mu$ decays by analyzing 2.92 fb$^{-1}$ data taken at 3.773 GeV.
From 9 hadronic decay modes of $D^-$ meson, the BESIII Collaboration accumulated $1703054\pm3405$ $D^-$ tags.
In this $D^-$ tag sample they observed $409.0\pm 21.2$ signal events for
$D^+ \rightarrow \mu^+\nu_\mu$ decays and measured the branching fraction
$B(D^+ \rightarrow \mu^+\nu_\mu)=(3.71 \pm 0.19 \pm 0.06)\times 10^{-4}$~\cite{BESIII_Dptomunu}.
They also measured $f_{D^+}|V_{cd}|=45.75\pm 1.20 \pm 0.39$ MeV and determined
$f_{D^+}=203.2 \pm 5.3 \pm 1.8$.
In addition, they extracted $|V_{cd}|=0.2210 \pm 0.0058 \pm 0.0047$~\cite{BESIII_Dptomunu,RongG_Charm2012}
from the leptonic $D^+$ decays for the first time.

    In 2013, the Belle Collaboration made new measurements of 
$D^+_s\rightarrow \mu^+\nu_\mu$ and
$D^+_s\rightarrow \tau^+\nu_\tau$ decays.
By analyzing 913 fb$^{-1}$ data collected near 10.6 GeV,
they measured 
$B(D^+_s\rightarrow \mu^+\nu_\mu)=(0.531 \pm 0.028 \pm 0.020)\%$,
$B(D^+_s\rightarrow \tau^+\nu_\tau)=(5.70 \pm 0.21^{+0.31}_{-0.30})\%$
and determined  $f_{D^+_s}=255.5 \pm 4.2 \pm 4.8 \pm 1.8$ MeV~\cite{Belle_Dstolnu}.

\subsection{Decay constants $f_{D^+_{(s)}}$}

    Historically the MARK-III, BES-I, BES-II, CLEO-c and BESIII Collaborations
studied the leptonic $D^+$ decays, but
only the BES-II, CLEO-c and BESIII Collaborations observed significant signal events
for $D^+\rightarrow \mu^+\nu_\mu$ decays,
and both the CLEO-c and BESIII Collaborations made precision measurements of this decay branching fraction.
Table~\ref{tab:B_DptoMunu}  summarizes $B(D^+\rightarrow \mu^+\nu_\mu)$
and some related quantities measured at the
CLEO-c~\cite{CLEO-c_Dptomunu} and BESIII~\cite{BESIII_Dptomunu} experiments.
The average of these two values of the measured branching fraction and related quantities~\cite{RongG_fpi0_Vcd_arXiv}
are also summarized in Table~\ref{tab:B_DptoMunu}.

\begin{table}[h]
  \centering
  \caption{
      Summary of $B(D^+\to\mu^+\nu_\mu)$, $f_{D^+}|V_{cd}|$, $f_{D^+}$ and $|V_{cd}|$.
          }
  \label{tab:B_DptoMunu}
\resizebox{\textwidth}{!}{
 \begin{tabular}{lcccc}
    \hline
    \hline
    {Quantity} & {$B(D^+\to\mu^+\nu_\mu)$ ($10^{-4}$)} & {$f_{D^+}|V_{cd}|$ (MeV) } & {$f_{D^+}$ (MeV)} & {$|V_{cd}|$ } \\
 \hline
    {CLEO-c~\cite{CLEO-c_Dptomunu}}  & {$3.82\pm 0.32\pm0.09$}       &  {N/A }   & 
                                                                         {$206.2\pm 8.6 \pm 2.6 $}   & {N/A }   \\
    {BESIII~\cite{BESIII_Dptomunu} }  & {$3.71\pm 0.19\pm0.06$} & {$45.75\pm 1.20 \pm 0.39$} & 
                                                                         {$203.2\pm 5.3 \pm 1.8$} & {$0.221\pm 0.006\pm 0.005$} \\ 
    {Average~\cite{RongG_fpi0_Vcd_arXiv} } & {$3.74\pm 0.17$} &  {$45.92\pm 1.04 \pm 0.15$} &
                                                                     {$203.9\pm 4.6 \pm 0.9$}  &  {$0.216\pm 0.005\pm 0.001$} \\
    \hline
    \hline
  \end{tabular}}
\end{table}

Table~\ref{tab:BF_Dstoln} summarizes branching fractions for $D^+_s\rightarrow \ell^+\nu_{\ell}$ decays
and relative branching ratios for $D_s^+\to\mu^+\nu_\mu$ decay measured at different experiments.
These measurements can be divided into three kinds of measurements, such as direct (D),
precision (P), and relative (R) measurements,
which are labeled as D, P and R in the last column of Table~\ref{tab:BF_Dstoln}, respectively.   
The relative measurements of the branching fraction for $D_s^+\to\mu^+\nu_\mu$ decay are determined
with $B(D^+_s \rightarrow \phi\pi^+)=(4.5\pm 0.4)\%$ quoted from PDG2014~\cite{pdg14}.

To determine the decay constants $f_{D^+_{(s)}}$ we separately consider all existing measurements or
some of these measurements of
branching fractions for $D^+_{(s)}\rightarrow \ell^+\nu_{\ell}$ decays.
Performing one free parameter, $f_{D_s^+}|V_{cs}|$, $\chi^2$ fit to
all (DPR) of these measured branching fractions with
inserting the well-measured $m_\mu=(105.6583715\pm0.0000035)$ MeV,
$m_\tau=(1776.82\pm0.16)$ MeV,
$m_{D_s^+}=(1968.30\pm0.11)$ MeV, and
$\tau_{D_s^+}=(500\pm7)\times10^{-15}$ s~\cite{pdg14} into Eq.(\ref{eq01})
yields the measured product $f_{D_s^+}|V_{cs}|=252.0\pm 3.7\pm 1.8$~\cite{FangY_fK0_Vcs_arXiv}.
Similarly, fitting both the direct (D) and the precision (P) measurements of these
branching fractions, or only fitting the precision (P) measurements of these
branching fractions yields corresponding values of $f_{D_s^+}|V_{cs}|$. 
These $f_{D_s^+}|V_{cs}|$ are all listed in Table~\ref{tab:fDs_DifMsrmnts}.
%
\begin{table}[h]
  \centering
  \caption{Summary of decay branching fractions $B(D_s^+\to\mu^+\nu_\mu)$, $B(D_s^+\to\tau^+\nu_\tau)$
           and branching ratio $R_B=B(D_s^+\to\mu^+\nu_\mu)/B(D_s^+\to\phi\pi^+)$ measured at different experiments.
           }
  \label{tab:BF_Dstoln}
  \resizebox{0.72\textwidth}{!}{
  \begin{tabular}{lccr}
    \hline
    \hline
    {\small Experiment} &               & {\small $B(D_s^+\to\mu^+\nu_\mu)$ (\%)}           &  Note \\
   {\small BES-I~\cite{BESI_leptonic} } &  &  {\small $1.5^{+1.3+0.3}_{-0.6-0.2}$} &  {\small D} \\
   {\small  ALEPH~\cite{aleph} } &  & {\small $0.68\pm0.11\pm0.18$}                &  {\small D} \\
   {\small  CLEO-c~\cite{CLEOc_munu} } & & {\small $0.565\pm0.045\pm0.017$}        &  {\small P} \\
   {\small  BaBar~\cite{BaBar_Dstolnu} } & & {\small $0.602\pm0.038\pm0.034$}      &  {\small P} \\
   {\small  Belle~\cite{Belle_Dstolnu} } & & {\small $0.531\pm0.028\pm0.020$}      &  {\small P} \\
    \hline
   {\small Experiment}                 & {\small $R_B$} &   $B(D_s^+\to\mu^+\nu_\mu)$ (\%)                                   & {\small R} \\
   {\small  BEATRICE~\cite{BEATRICE} } & {\small $0.23\pm0.06\pm0.04$}        & {\small $1.04\pm  0.27\pm 0.18\pm0.09$}    & {\small R} \\
   {\small  CLEO-II~\cite{CLEOII_munu} } &  {\small $0.173\pm0.023\pm0.035$}  & {\small $0.779\pm 0.104\pm 0.158\pm0.069$} & {\small R} \\
   { BaBar~\cite{BaBar_munu} } & {\small $0.143\pm0.018\pm0.006$}             & {\small $0.644\pm 0.081\pm 0.027\pm0.057$} & {\small R} \\
    \hline
   {\small Experiment} & & {\small $B(D_s^+\to\tau^+\nu_\tau)$ (\%)}          &          \\
   {\small  L3~\cite{L3} } & & {\small $7.4\pm2.8\pm2.4$}            &  {\small D}  \\
   {\small  OPAL~\cite{opal} } & & {\small $7.0\pm2.1\pm2.0$ }       &  {\small D} \\
   {\small  ALEPH~\cite{aleph} } & & {\small $5.79\pm0.77\pm1.84$}   &  {\small D} \\
   {\small  CLEO-c~\cite{CLEOc_taunu} } & & {\small $5.58\pm0.33\pm0.13$ } & {\small P}  \\
   {\small  BaBar~\cite{BaBar_Dstolnu} } & & {\small $5.00\pm0.35\pm0.49$ }  & {\small P} \\
   {\small  Belle~\cite{Belle_Dstolnu} } & & {\small $5.70\pm0.21^{+0.31}_{-0.30}$ }  & {\small P}  \\
    \hline
    \hline
  \end{tabular}}
\end{table}
%

\begin{table}[h]
  \centering
  \caption{
      $f_{D^+_s}|V_{cs}|$ and $f_{D^+_s}$ determined by fitting different decay branching fractions in groups of
      DPR, DP and P shown in Table~\ref{tab:BF_Dstoln}.
          }
  \label{tab:fDs_DifMsrmnts}
  \begin{tabular}{lccr}
    \hline
    \hline
    {\small Group of $B(D^+_s\rightarrow \ell^+\nu_{ell})$}   & {\small DPR } & {\small DP} & {\small P}  \\
 \hline
   {\small $f_{D^+_s}|V_{cs}|$ } & {\small $252.0\pm 3.7\pm 1.8$~\cite{FangY_fK0_Vcs_arXiv}} & {\small $250.7\pm 3.8\pm 1.8$} & 
                                                                                                       {\small $250.2\pm 3.8\pm 1.8$} \\
   {\small $f_{D^+_s}$ }            & {\small $258.9\pm 4.2$~\cite{FangY_fK0_Vcs_arXiv}} & {\small $257.5\pm 4.3$} &
                                                                                                  {\small $257.0\pm 4.3$} \\
   \hline \hline
  \end{tabular}
\end{table}

\begin{table}[h]
  \centering
  \caption{
      Summary of
   $f_{D^+_s}/f_{D^+}$ determined with the values for $f_{D^+_s}$ shown in Table~\ref{tab:fDs_DifMsrmnts} 
   and the average value for $f_{D^+}$ determined 
   from $B(D^+\rightarrow \mu^+\nu_\mu)$ measured at the BESIII and CELO-c experiments.
          }
  \label{tab:fDs_o_fD}
  \begin{tabular}{lccr}
    \hline
    \hline
    {\small Group of $B(D^+_s\rightarrow l^+\nu_l)$}  & {\small DPR } & {\small DP} & {\small P}  \\
 \hline
     {\small $f_{D^+_s}/f_{D^+}$ }  &   $1.270\pm 0.036$   &  $1.263\pm 0.036$    &  $1.260\pm 0.036$   \\
    \hline
    \hline
  \end{tabular}
\end{table}

Dividing $f_{D^+}|V_{cd}|=(45.92 \pm 1.04 \pm 0.15)~\rm MeV$~\cite{RongG_fpi0_Vcd_arXiv} 
by $|V_{cd}|=0.22522\pm0.00061$ from global fit in the SM~\cite{pdg14}
yields~\cite{RongG_fpi0_Vcd_arXiv}
$$f_{D^+}=(203.9\pm 4.6 \pm 0.9)=(203.9\pm 4.7)~~\rm MeV,$$
which is obtained from both the BESIII~\cite{BESIII_Dptomunu} 
and CLEO-c~\cite{CLEO-c_Dptomunu} measurements of branching fraction for $D^+\to\mu^+\nu_\mu$ decay.
Similarly, dividing $f_{D_s^+}|V_{cs}|=(252.0 \pm 3.7 \pm 1.8)~\rm MeV$ 
obtained by fitting all decay branching fractions listed in Table~\ref{tab:BF_Dstoln} 
by $|V_{cs}|=0.97343\pm0.00015$ from global fit in the SM~\cite{pdg14}
yields~\cite{FangY_fK0_Vcs_arXiv}
$$f_{D_s^+}=(258.9\pm 3.8\pm 1.8)=(258.9\pm 4.2)~~\rm MeV.$$
As a comparison Table~\ref{tab:fDs_DifMsrmnts} lists this $f_{D_s^+}$ and other values of $f_{D_s^+}$ determined by separately fitting 
the branching fractions in group DP and in group P,
where the statistical and systematic errors are combined together.
With these determined values of $f_{D_s^+}$ and $f_{D^+}$ we find 
$$f_{D_s^+}/f_{D^+}=1.270\pm 0.036,$$
which is $2.8\sigma$ ($\sigma$ is standard deviation) larger
than the most precision value $[f_{D^+_s}/f_{D^+}]_{\rm LQCD} =1.171\pm 0.001\pm0.003$ 
calculated in LQCD~\cite{arXiv_1407_3772_LQCD_fDs_fD}.
Table~\ref{tab:fDs_o_fD} compares this determined ratio and other values of the ratio 
for which the values of $f_{D^+_s}$ are determined by fitting the branching fractions
in group DP and in group P.
The last two ratios shown in Table~\ref{tab:fDs_o_fD} are larger than the predicted ratio 
calculated in LQCD~\cite{arXiv_1407_3772_LQCD_fDs_fD}
by $2.6\sigma$ and $2.5\sigma$, respectively.

\section{Recent results on semileptonic $D$ Decays}
   Recently, the BESIII Collaboration reported preliminary results on
measurements of $D^0\to \pi^-e^+\nu_e$ and
$D^0\to K^-e^+\nu_e$ decays. From 2.92 fb$^{-1}$ data taken at 3.773 GeV,
the BESIII Collaboration accumulated $(279.3\pm0.4)\times 10^4$ $\bar D^0$ tags 
with five hadronic decay modes.
In this sample of $\bar D^0$ tags, they observed 
$6297\pm87$ and
$70727\pm278$
signal events for 
$D^0\to \pi^-e^+\nu_e$ and $D^0\to K^-e^+\nu_e$
decays, respectively, and measured the branching fractions
$B(D^0\to \pi^-e^+\nu_e)=(0.2950\pm0.0041\pm0.0026)\%$ and 
$B(D^0\to K^-e^+\nu_e)=(3.505\pm0.014\pm0.033)\%$~\cite{BESIII_D0Kenu}. 
By analyzing differential rates of these two decays, 
they directly measured
$f_+^\pi(0)|V_{cd}|=0.1420\pm0.0024\pm0.0010$ and
$f_+^K(0)|V_{cs}|=0.7196\pm0.0035\pm0.0041$~\cite{BESIII_D0Kenu}.

    The BaBar Collaboration recently reported a measurement of $D^0\to \pi^-e^+\nu_e$ decays.
From a data sample taken near 10.6 GeV,
they observed 5300 $D^0\to \pi^-e^+\nu_e$ decays and measured a relative
branching ratio $R_B = B(D^0\to \pi^-e^+\nu_e)/B(D^0\to K^-\pi^+)=0.0702\pm 0.0017\pm 0.0023$
and determined $B(D^0\to \pi^-e^+\nu_e)=(0.2770 \pm 0.0068 \pm 0.0092\pm 0.0037)\%$. 
They also measured  $f_+^\pi(0)|V_{cd}|=0.1374\pm0.0038\pm0.0022\pm0.0009$~\cite{BaBar_D0topienu}.
Multiplying their measured $f^K_+(0)=0.727\pm0.007\pm0.005\pm0.007$~\cite{BaBar_D0toKenu}
by $|V_{cs}|=0.9729\pm0.0003$ used in their paper yields $f_+^K(0)|V_{cs}|=0.707\pm0.007\pm0.005\pm0.007$.

    In 2008, the CLEO-c Collaboration studied the semileptonic decays of $D^0\to \pi^-e^+\nu_e$, $D^0\to K^-e^+\nu_e$,
$D^+\to\pi^0e^+\nu_e$ and $D^+\to\bar K^0e^+\nu_e$ by analyzing 818 pb$^{-1}$ data taken at 3.773 GeV.
They measured $f_+^\pi(0)|V_{cd}|=0.150\pm0.004\pm0.001$ and
$f_+^K(0)|V_{cs}|=0.719\pm0.006\pm0.005$~\cite{CLEOc_DtoPenu}.

  The Belle Collaboration made measurements of $D^0\to \pi^-e^+\nu_e$ and $D^0\to K^-e^+\nu_e$
decays in 2006. By analyzing 282 fb$^{-1}$ data taken at 10.58 GeV they measured the form factors
$f_+^\pi(q^2)$ and $f_+^K(q^2)$, and determined $f_+^{\pi}(0)=0.624\pm 0.020 \pm 0.030$
as well as  $f_+^K(0)=0.695\pm 0.007 \pm 0.022$~\cite{Belle_PRL97_061804_y2006}.

Figure 1 shows comparison of these measured form factors along with theoretical predictions for these two form factors.
\begin{figure}[htb]
\centering
\includegraphics[height=1.7in,width=2.9in]{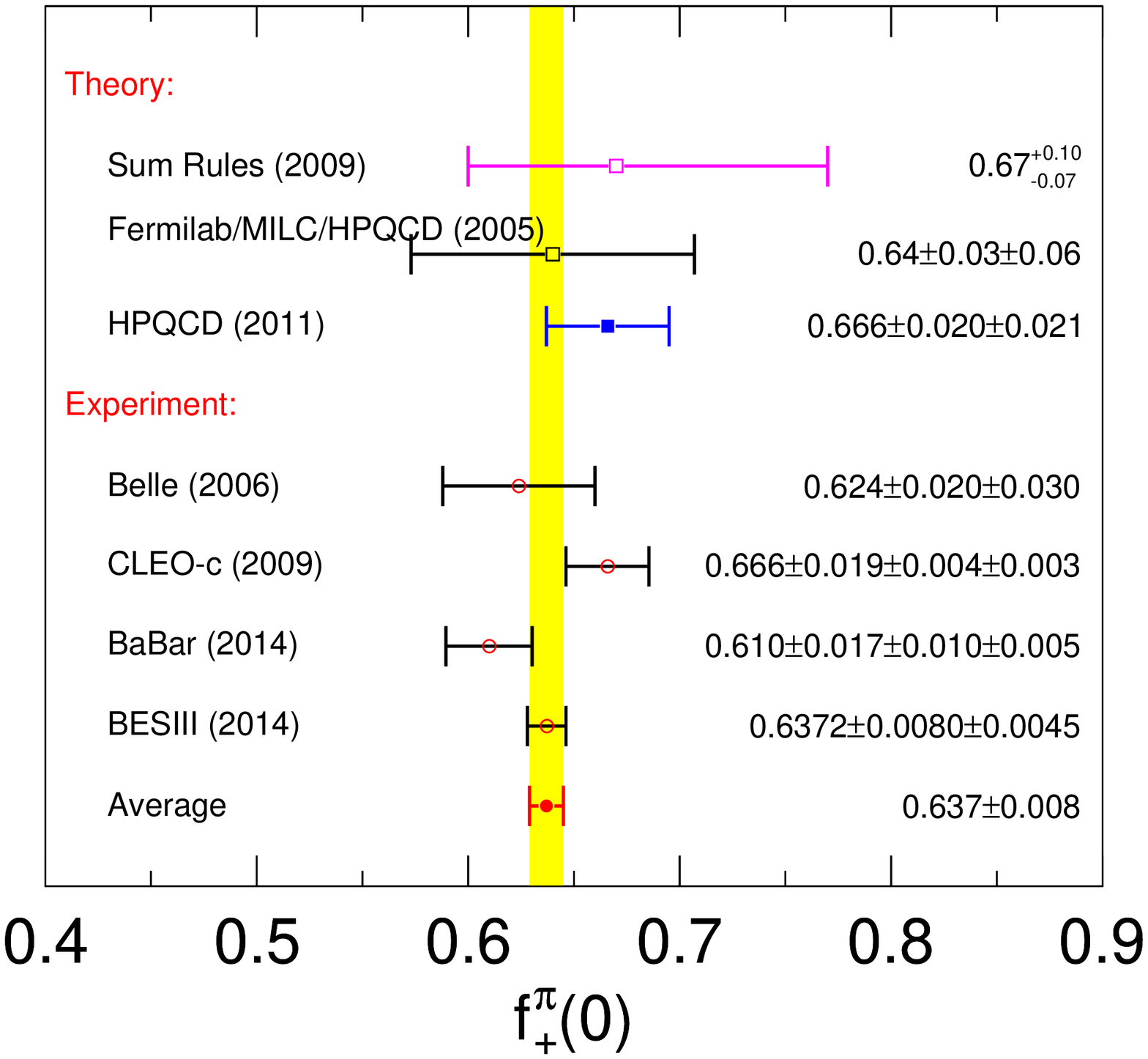}
\includegraphics[height=1.7in,width=2.9in]{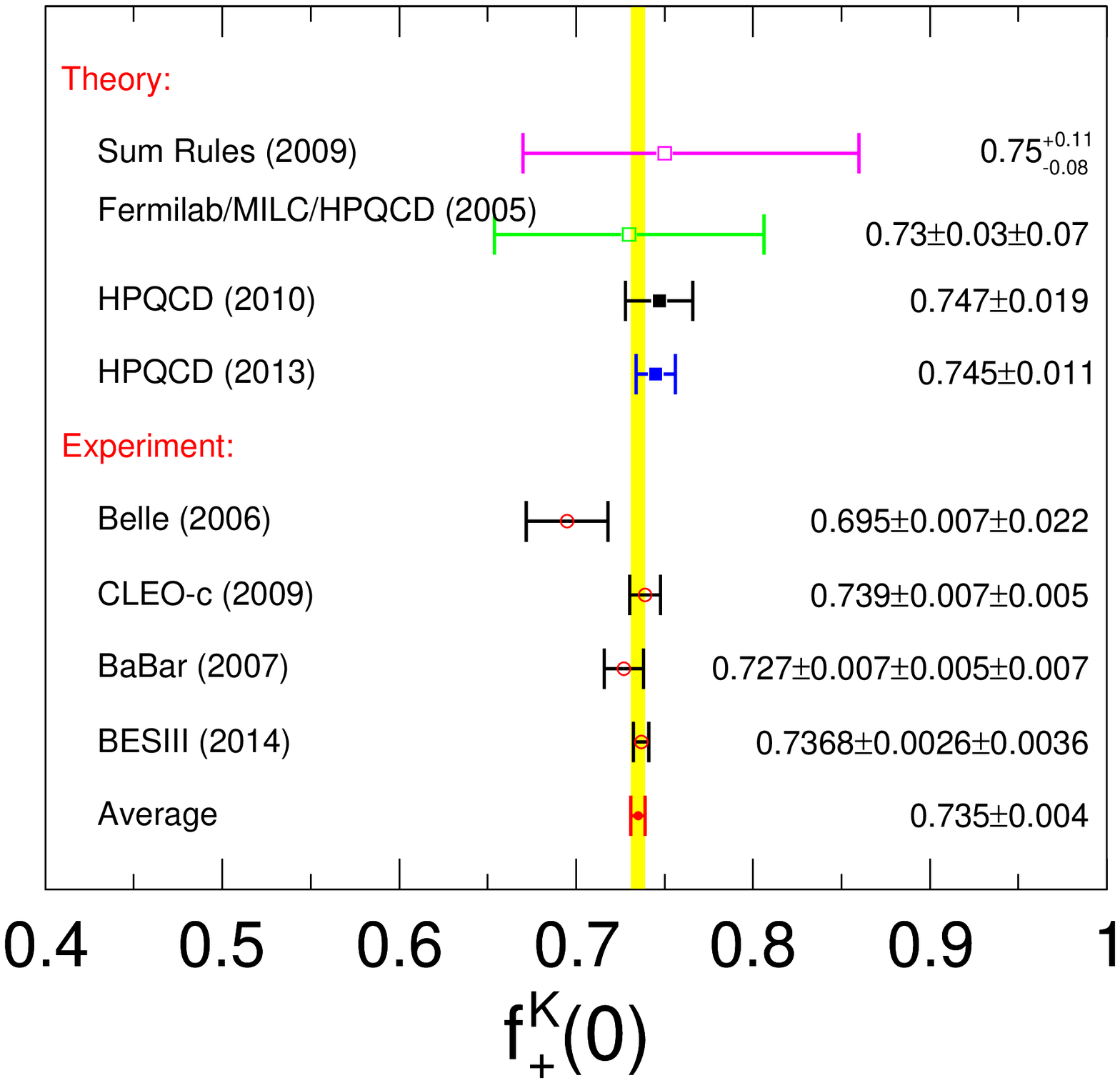}
\vspace{-0.3cm}
\caption{Comparison of measured form factors along with theoretical predictions for these.}
\label{fig:magnet}
\end{figure}

Combining three measurements of $f_+^{\pi(K)}(0)|V_{cd(s)}|$ from the BaBar, BESIII and CLEO-c,
we obtain the average~\cite{RFM_arXiv:1409.1068}
$$f^\pi_+(0)|V_{cd}| = 0.143\pm0.002,~~~~f^K_+(0)|V_{cs}| = 0.718\pm0.004,$$
where the errors are the combined statistical and systematic errors together.

\section{Extractions of $|V_{cd}|$ and $|V_{cs}|$}

Using the determined $f_{D^+}|V_{cd}|=45.92\pm 1.04 \pm 0.15$ MeV shown in Table~\ref{tab:B_DptoMunu} 
and $f_{D^+_s}|V_{cs}|=250.2\pm 3.8 \pm 1.8$ MeV shown in Table~\ref{tab:fDs_DifMsrmnts},
in conjunction with the averaged values $f_{D^+}=(209.2\pm 3.3)$ MeV and
$f_{D^+_s}=(248.6\pm 2.7)$ MeV from the Flavor Lattice Averaging Group, we find
the CKM matrix elements $|V_{cd}|^{D^+ \rightarrow \mu^+\nu_\mu}=0.219 \pm 0.005_{\rm exp} \pm 0.004_{\rm LQCD}$
and $|V_{cs}|^{D^+_s \rightarrow \ell^+\nu_{\ell}}=1.006 \pm 0.016_{\rm exp} \pm 0.013_{\rm LQCD}$.
Similarly, with the measured 
$f_+^{\pi}(0)|V_{cd}|=0.143\pm0.002$ and $f_+^{K}(0)|V_{cs}|=0.718\pm0.004$
together with $[f_+^\pi(0)]_{\rm LQCD}=0.666\pm 0.029$~\cite{HPQCD_fpi0} 
and $[f_+^K(0)]_{\rm LQCD}=0.747\pm 0.019$~\cite{LQCD_fK}
we extract the CKM matrix elements 
$|V_{cd}|^{D \rightarrow \pi e^+\nu_e} = 0.215 \pm 0.003_{\rm exp} \pm0.009_{\rm LQCD}$ 
and $|V_{cs}|^{D \rightarrow \pi e^+\nu_e}=0.961 \pm 0.005_{\rm exp} \pm0.024_{\rm LQCD}$.
Combining these $|V_{cd(s)}|^{D\rightarrow \pi(K) e\nu_e}$ and $|V_{cd(s)}|^{D^+_{(s)} \rightarrow l^+\nu_l}$
together, we find~\cite{RFM_arXiv:1409.1068} 
$$|V_{cd}| =0.218 \pm 0.005,~~~~~~|V_{cs}| =0.985 \pm 0.015.$$   
The above $|V_{cd}|$ and $|V_{cs}|$ are extracted from measurements of leptonic $D^+_{(s)}$ decays
and semi-leptonic $D$ decays performed at the BaBar, Belle, BESIII and CLOE-c experiments.
In these measurements of semi-leptonic $D$ decays, the $f_+^{\pi(K)}(0)|V_{cd(s)}|$ were all measured
by analyzing differential rates of $D \rightarrow \pi(K)e^+\nu_e$ decays.

Beyond these four measurements of differential decay rates, many other experiments measured some other
quantities rather than the differential decay rates. In order to reduce the experimental uncertainties in 
$|V_{cd}|$ and $|V_{cs}|$, several authors analyzed all existing measurements of
these leptonic $D^+_{(s)}$ and semi-leptonic $D$ decays to extract $|V_{cd}|$ and $|V_{cs}|$.
By globally analyzing all existing measurements together with the recent most precision values for 
these form factors and decay constants calculated in LQCD,
they extracted~\cite{RongG_fpi0_Vcd_arXiv, FangY_fK0_Vcs_arXiv}
$$|V_{cd}| =0.2157 \pm 0.0045,~~~~~~~|V_{cs}| =0.983 \pm 0.011,$$
which are the most precise determinations of $|V_{cd}|$ and $|V_{cs}|$ up to date.
Figure 2 shows comparison of these extracted $|V_{cd}|$ and $|V_{cs}|$ 
as well as other ones given by the HFAG~\footnote{Personal communication with A. Zupanc.} along with the values from PDG2014.
\begin{figure}[htb]
\centering
\includegraphics[height=1.7in,width=2.9in]{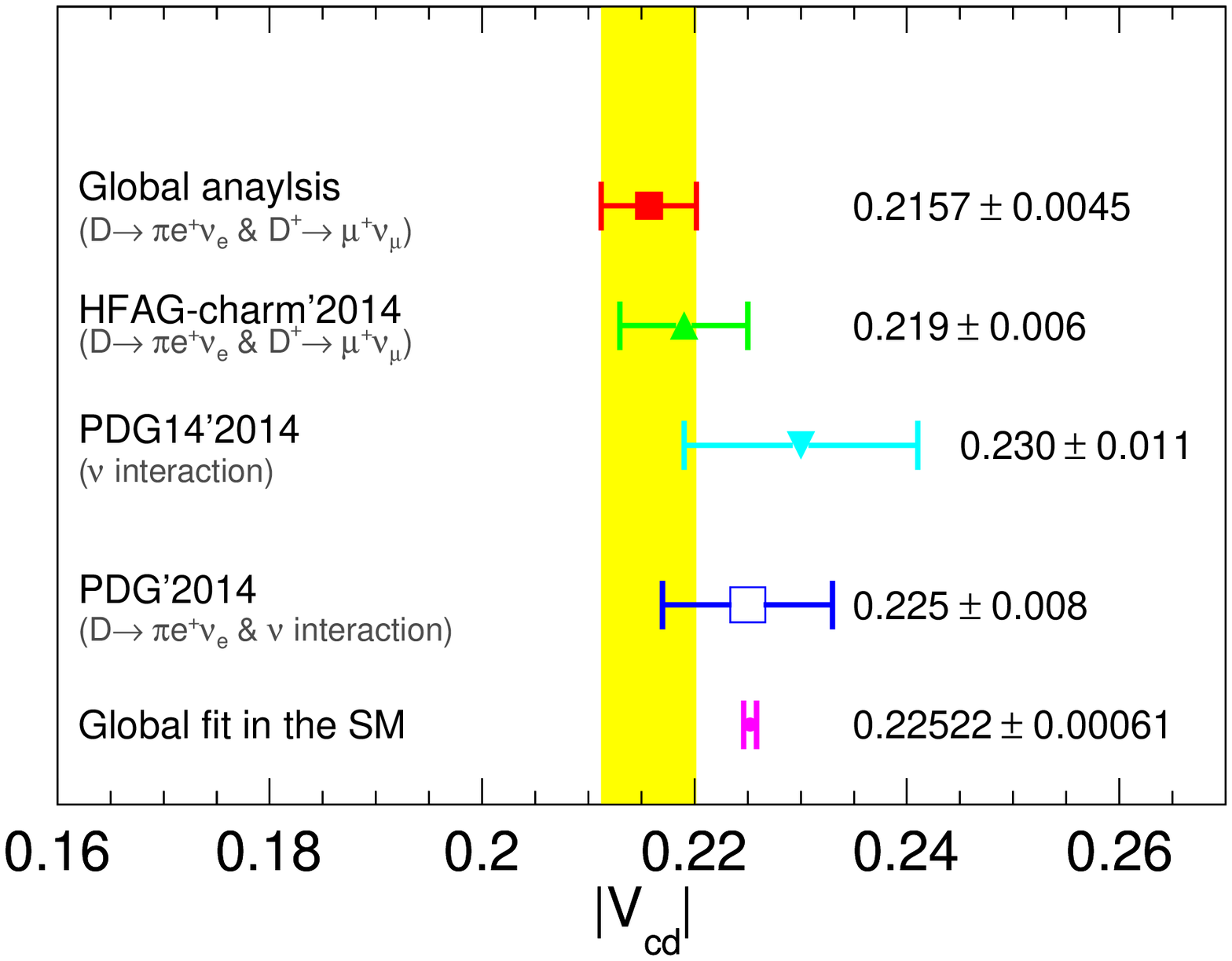}
\includegraphics[height=1.7in,width=2.9in]{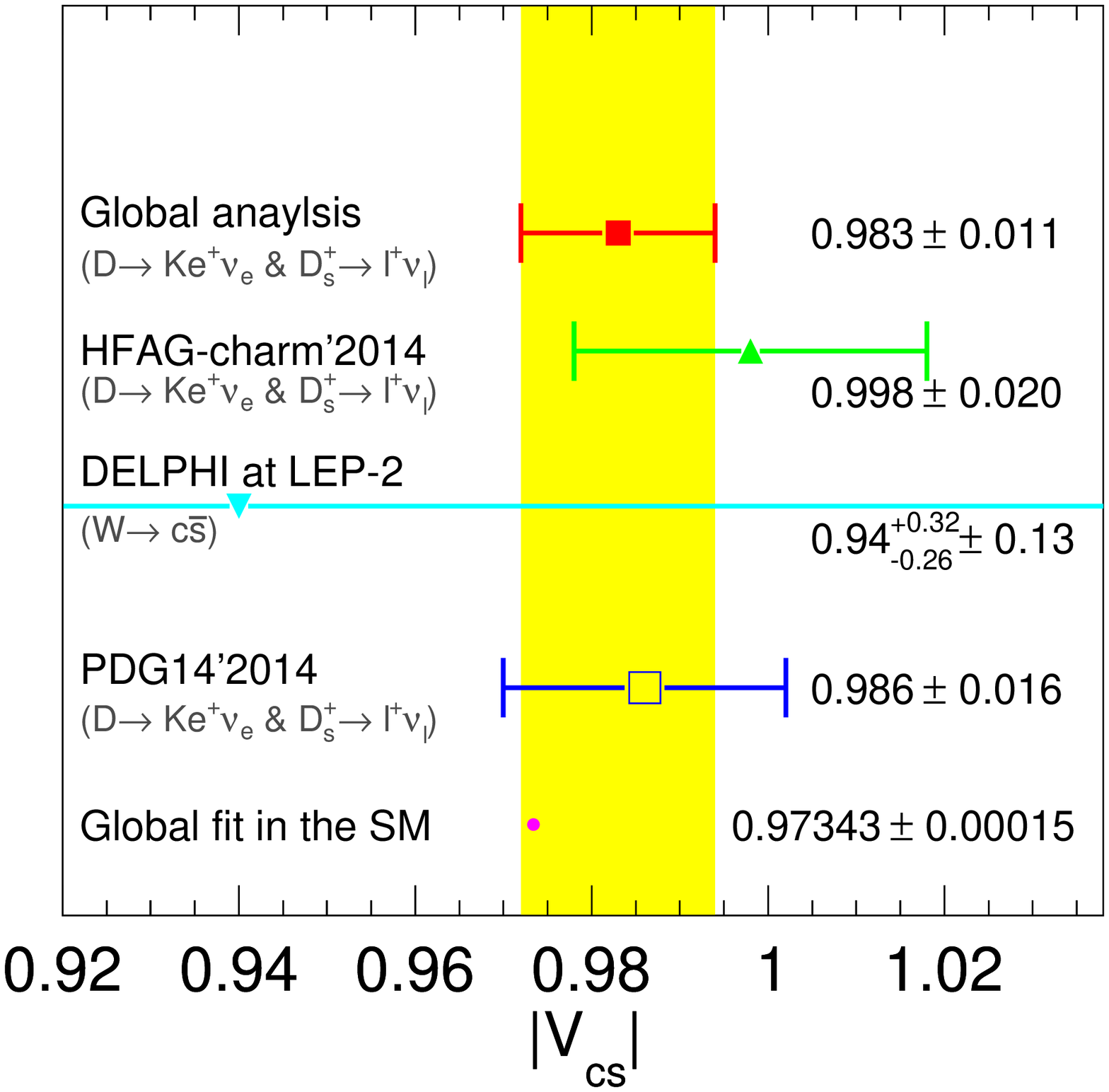}
\vspace{-0.3cm}
\caption{Comparison of the extracted $|V_{cd}|$ and $|V_{cs}|$ along with PDG2014 values.}
\label{fig:magnet}
\end{figure}

\section{Summary}
    Analyzing all existing measurements of leptonic $D^+_{(s)}$ decays
measured at different experiments yields 
$f_{D^+}=203.9\pm 4.7$ MeV, $f_{D^+_s}=258.9\pm 4.2$ MeV, 
and $f_{D^+_s}/f_{D^+}=1.270\pm 0.036$ which is larger than 
the ratio calculated in LQCD~\cite{arXiv_1407_3772_LQCD_fDs_fD}  
by $2.8\sigma$. If only analyzing the branching fractions for $D^+_s \rightarrow \ell^+\nu_{\ell}$ decays 
measured at the BaBar, Belle and CELO-c experiments, these become to be  
$f_{D^+_s}=257.0\pm 4.4$ MeV, and $f_{D^+_s}/f_{D^+}=1.260\pm 0.036$, which is
larger than the predicted ratio by $2.5\sigma$.
The average of the measured $f_+^{\pi(K)}(0)$ from the BaBar, Belle, BESIII and CLEO-c experiments
are consistent within error with
those calculated in LQCD, but with a factor of 3 (4) more precisions than
that calculated in LQCD.
Analyzing all measurements of the leptonic $D^+_{(s)}$ and semi-leptonic $D$ 
decays together, in conjunction with the recent LQCD calculations for these form factors
and decay constants, ones find $|V_{cd}| =0.2157 \pm 0.0045$
and $|V_{cs}| =0.983 \pm 0.011$,
which are the most precise determinations of $|V_{cd}|$ and $|V_{cs}|$ up to date.
At present, the uncertainties of
the extracted $|V_{cd}|$ and $|V_{cs}|$ are still dominated 
by the error of $f_+^{\pi(K)}(0)$ calculated in LQCD.
If these errors of $f_+^{\pi(K)}(0)$ calculated in LQCD could be negligible, the relative accuracy 
of $|V_{cd}|$ and $|V_{cs}|$ from semi-leptonic $D$ decays could reach to
$\Delta |V_{cd}|/|V_{cd}|\sim 1.2\%$ and $\Delta |V_{cs}|/|V_{cs}|\sim 0.5\%$.

\end{document}